\documentclass{CHEP2006}
\usepackage{graphicx}

\newcommand{\CClass}[1]{\emph{#1}}

\newcommand{\PaxFV}{\CClass{PaxFourVector}}
\newcommand{\PaxVX}{\CClass{PaxVertex}}
\newcommand{\PaxCO}{\CClass{PaxCollision}}

\newcommand{\PaxEI}{\CClass{PaxEventInterpret}}

\newcommand{\PaxP}{\CClass{PaxProcess}}
\newcommand{\PaxAP}{\CClass{PaxAutoProcess}}
\newcommand{\PaxPF}{\CClass{PaxProcessFactory}}
\newcommand{\PaxEITT}{\CClass{PaxEventInterpretTTree}}

\begin{document}
\title{Concepts, Developments and Advanced Applications of the PAX Toolkit}

\author{
S. Kappler\thanks{Corresponding author: Steffen.Kappler@cern.ch}, M. Erdmann, M. Kirsch, G. M\"uller (RWTH Aachen university, Germany), \\
J. Weng (CERN, Geneva, Switzerland), 
A. Flo\ss dorf (DESY, Hamburg, Germany), \\
U. Felzmann, G. Quast, C. Saout, A. Schmidt (Karlsruhe university, Germany)
}

\maketitle

\begin{abstract}
The Physics Analysis eXpert (PAX) is an open source toolkit for high energy physics analysis.
The C++ class collection provided by PAX is deployed in a number of analyses with complex event 
topologies at Tevatron and LHC. In this article, we summarize basic concepts and class structure 
of the PAX kernel. We report about the most recent developments of the kernel and introduce two 
new PAX accessories. The PaxFactory, that provides a class collection to facilitate event hypothesis
evolution, and VisualPax, a Graphical User Interface for PAX objects.
\end{abstract}

\section{Introduction}
Physics analyses at modern collider experiments enter a new dimension of event complexity. 
At the LHC, for instance, physics events will consist of the final state products of the 
order of 20 simultaneous collisions. In addition, a number of today's physics questions 
is studied in channels with complex event topologies and configuration ambiguities occurring 
during event analysis. 

The Physics Analysis eXpert toolkit (PAX) is a continuously maintained and advanced C++ class collection, 
specially designed to assist physicists in the analysis of complex scattering processes \cite{PAX02,PAX03,PAX04,PAX06}.
PAX is realized in the C++ programming language \cite{CPPSTL}.
It provides additional functionality in top of the vector algebra of the 
widely-spread libraries CLHEP \cite{CLHEP} (default) or ROOT \cite{ROOT}.
The PAX container model as well as file I/O are based on the C++ 
Standard Template Library (STL) \cite{CPPSTL}. 

\section{The PAX kernel}

The class collection of the PAX kernel allows the definition of an abstraction layer beyond 
detector reconstruction by providing a generalized, persistent HEP event container with three types 
of physics objects (particles, vertices and collisions), relation management and file I/O scheme. 
The PAX event container is capable of storing the complete information of multi-collision 
events (including decay trees with spatial vertex information, four-momenta as well as 
additional reconstruction data). An automated copy functionality for the event container 
allows the analyst to consistently duplicate event containers for hypothesis evolution, 
including its physics objects and relations. PAX physics objects can hold pointers to an 
arbitrary number of instances of arbitrary C++ classes, allowing the analyst to keep track 
of the data origin within the detector reconstruction software. Further advantages arising 
from the usage of the PAX toolkit are a unified data model and nomenclature, and therefore 
increased code lucidity and more efficient team work. The application of the generalized 
event container provides desirable side-effects, such as protection of the physics analysis 
code from changes in the underlying software packages and avoidance of code duplication by 
the possibility of applying the same analysis code to various levels of input data.

\subsection{PAX physics objects}
The three types of generalized physics objects provided by the PAX kernel are:
\begin{itemize}
\item{Particles (or reconstructed objects), i.e.\ Lorentz-vectors, are
      represented by the class \PaxFV.}
\item{Vertices, i.e.\ three-vectors,
      represented by the class \PaxVX, are foreseen to realize particle 
      decays.}
\item{Collisions, represented by the class \PaxCO, are foreseen to allow the
      separation of multiple interactions in high-luminostity environments.}
\end{itemize}
The vector characteristics of the classes \PaxFV\ and \PaxVX\ is inherited from 
the corresponding classes of the CLHEP or ROOT libraries. 
Commonly needed, additional properties such as a name, particle-id, status, 
charge, a workflag etc.\ can be stored in data members. Specific information 
complementary to data members, such as b-tags, jet cone sizes or energy 
corrections, for instance, can be stored in the so-called user records (i.e.\ 
collections of string-double pairs). 

Each PAX physics object can record pointers to an arbitrary 
number of instances of arbitrary C++ classes. This way, the 
user can keep track of the data origin within the detector 
reconstruction software, for instance. Access to the 
pointers is possible at the same runtime
during any later stage of the analysis. 

Copy constructors are provided to perform deep copies of PAX physics objects. 
When copying a PAX physics object, all pointers are copied as well. 
For the convenience of reconstructing particle decay chains, 
PAX physics objects are enabled to establish relations and can be organized in 
containers based on the STL class templates \CClass{map$<$key, item$>$} and
\CClass{multimap$<$key, item$>$}, respectively. 

\subsection{Event container}

PAX provides a generalized event container for storage and handling of the 
complete information of one multicollision event including 
decay trees, spatial vertex information, four-momenta as well as additional 
reconstruction data in the user records.  

This container is represented by the class \PaxEI. This class is so named, because 
it is intended to represent a distinct interpretation of an event configuration 
(e.g.\ connecting particles to the decay tree according to one out of a
number of hypotheses, applying different jet energy corrections, etc.). 
To facilitate the development of numerous parallel or subsequent event 
interpretations, the \PaxEI\ class features a copy constructor, which 
provides a deep copy of the event container with all data members, 
physics objects, and their redirected relations.

The PAX toolkit offers a file I/O scheme for persistent
storage of the event container, based on STL streams. 
It allows the user to write the contents of 
\PaxEI\ instances with all contained physics objects
as well as their relations to PAX data files. 
When restoring the data from file, an empty \PaxEI\ instance is filled 
with the stored data and objects and all object relations are reproduced.

The PAX data file format is multi-version and multi-platform compatible 
and consists of binary data chunks, that allow file structure checks and 
fast positioning. 

PAX also provides the possibility to write/read \PaxEI\ instances to/from 
strings. This way, the user can store PAX objects to any 
data format supporting strings or binary data fields like databases or 
experiment specific data formats.

All classes of the PAX kernel can be used in compiled mode within the 
C-interpreter of ROOT. A more detailed description of the PAX kernel 
functionalities can be found in reference \cite{PAX06}. 

\section{The PAX accessories}
\subsection{The PAX factory}

The PaxFactory is an accessory to the PAX kernel that facilitates the bookkeeping 
of event hypothesis evolution. 

\begin{figure}[htb]
\centering
\includegraphics*[width=65mm]{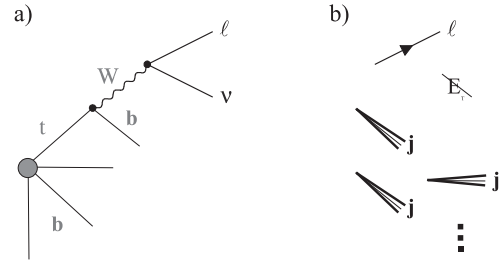}
\caption{a) Schematic view of single top production with leptonic top decay.
         b) The visible reconstructed partons of this channel.}
\label{fig_ambiguities.eps}
\end{figure}

The class \PaxP\ is designed to allow the evolution 
of different combinatorial hypotheses (\emph{event interpretations}) of an event 
according to a certain physics process. This task arises during the a priori 
ambiguous partonic reconstruction of processes with multiple reconstructed objects 
of the same type. Figure \ref{fig_ambiguities.eps} gives single top 
production with leptonic top decay for illustration: permuting the plotted three jets 
from figure \ref{fig_ambiguities.eps}.b at the t$\rightarrow$Wb vertex in figure 
\ref{fig_ambiguities.eps}.a provides three different hypotheses. In addition, the 
normally two-fold ambiguity of the longitudinal neutrino momentum, as provided by 
a W-mass constraint, doubles the number of combinatorial hypothesis for the t-quark. 
With the \PaxP\ class, the analyst can store and manage an arbitrary number of 
event-interpretations including their physics objects and relations. Like all other 
PAX objects, a \PaxP\ instance allows to store data in the user records and can record 
pointers to an arbitrary number of instances of arbitrary C++ classes. 

A higher degree of automation at the evolution of combinatorial hypotheses is provided 
by the class \PaxAP. This derivative of the \PaxP\ class features automatic evolution
of all possible combinatorial hypotheses of an event. The rules, according to which
these hypotheses are evolved, are defined by user-defined static process-model, that
is passed to the \PaxAP\ instance at construction time. 
This process-model is a \PaxEI\ instance containing a prototype of the process 
decay-chain with parameters customizing the behaviour of the \PaxAP\ class. 
The remaining step, to be done by the analyst, is to further process the evolved
event interpretations in standard decision techniques, for instance. 

\begin{figure}[htb]
\centering
\includegraphics*[width=65mm]{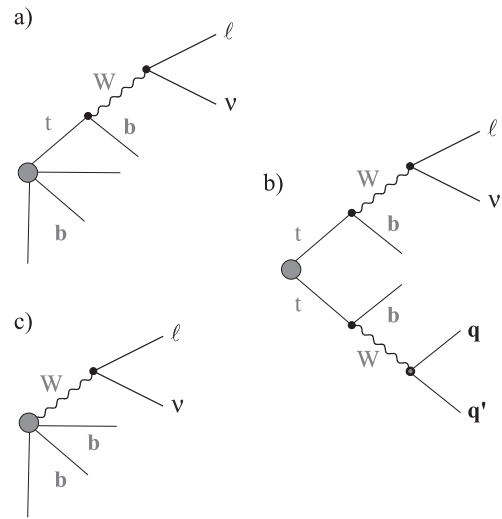}
\caption{Schematic view of a) single top production 
         and its main backgrounds b) top-pair production 
                              and c) jet-associated W production.}
\label{fig_processes.eps}
\end{figure}

A further aspect of hypothesis evolution, besides the resolution of combinatorial ambiguities, 
is the parallel evolution of different physics process hypotheses of an event. As illustrated in 
figure \ref{fig_processes.eps} for the analysis of single top production,
the analyst might want to distinguish the signal channel (figure \ref{fig_processes.eps}.a) 
from its main backgrounds individually (figure \ref{fig_processes.eps}.b and \ref{fig_processes.eps}.c).

To allow easy management of different physics process hypotheses of an event, the 
The class \PaxPF\ provides storage and easy access to an arbitrary number of processes 
(i.e. instances of the class \PaxP\ or derivatives) as well as user records 
and recording of arbitrary C++ pointers.

While the \PaxEI\ instances with their physics objects are intented to 
remain in memory for one event, the classes of the PaxFactory accessory 
are designed for lifetimes of up to one computing  job. Therefore, the classes 
\PaxP\ and \PaxPF\ provide virtual member functions to be called 
at the beginning and end of a job, at the beginning and end of a run, 
and, of course, at event analysis and event finishing time. By default, the 
methods of the \PaxPF\ class invoke the corresponding methods of all managed 
\PaxP\ instances.

The PaxFactory accessory provides the class \PaxEITT, which allows to automatically 
copy selected observables from \PaxEI\ instances on an event-by-event basis into the 
TTree of a ROOT file. Those observables may be kinematic data or user records of 
certain contained physics objects as well as user-records of the event-interpretation.
The automatic copy is performed according to a user-defined, static copy-model in the
form of a \PaxEI\ instance, that is passed to the \PaxEITT\ instance at construction time. 

\begin{figure}[htb]
\centering
\includegraphics*[width=80mm]{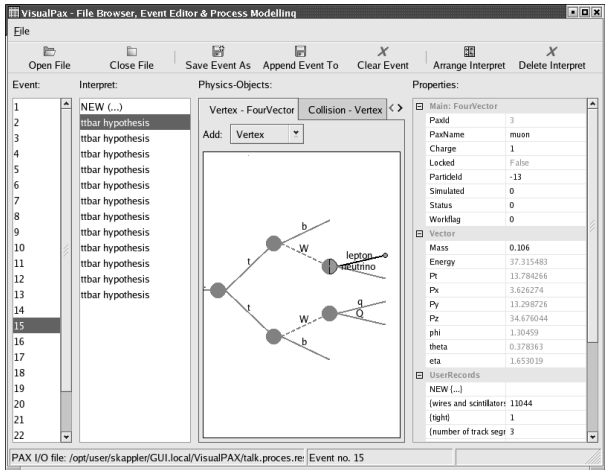}
\caption{The Graphical User-Interface of VisualPax allows browsing of PAX I/O files
and editing of \PaxEI\ instances.}
\label{fig_visualpax.eps}
\end{figure}

\begin{figure}[htb]
\centering
\includegraphics*[width=80mm]{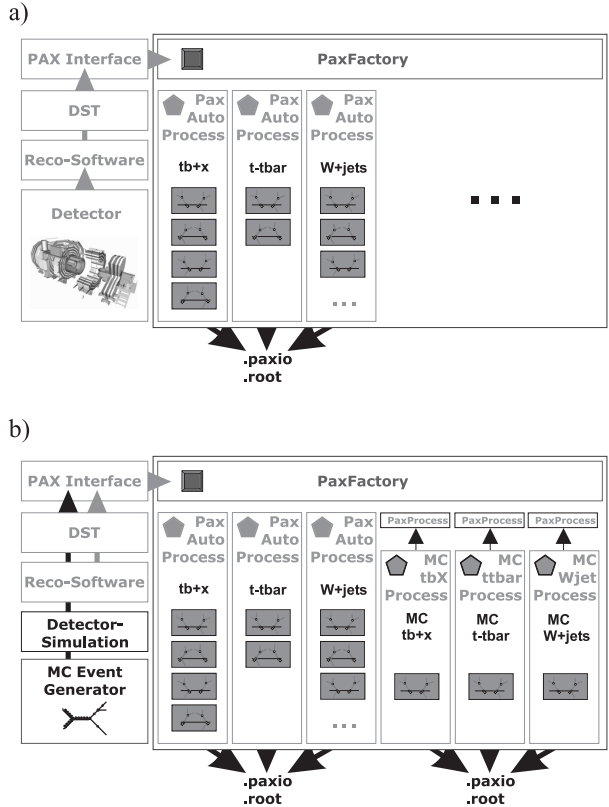}
\caption{Example scheme for the use of the PAX kernel plus accessories in advanced physics analyses.
         The same software is used in different configurations when running over a) experiment data 
         and b) Monte-Carlo data.}
\label{fig_factory.eps}
\end{figure}

\subsection{Visual PAX}

VisualPax is a recenlty developed accessory to the PAX kernel that allows browsing of PAX I/O files
and editing of \PaxEI\ instances in a Graphical User-Interface.
VisualPax is based on the wxWidgets open source, cross-platform native user-interface framework \cite{wxWidgets}. 
As shown in figure \ref{fig_visualpax.eps}, VisualPax allows to graphically display and modify 
event interpretations including properties and decay chains of the contained physics objects. 
Therefore, with the help of VisualPax and PAX I/O files, the 
process-models for \PaxAP\ instances as well as the copy-models for \PaxEITT\ instances can be 
managed in a comfortable way.

\section{PAX advanced analysis example}

Advanced physics analyses realized with the PAX kernel and accessories can be designed according to 
the schema shown in figure \ref{fig_factory.eps}.

When running over experiment data, a dedicated, experiment-specific interface class 
for filling the PAX containers (i.e.\ \PaxEI\ instances) represents the 
interface between detector reconstruction software and the PAX factory. 
Once all relevant information is filled, the objects are passed to a \PaxPF\ instance 
that manages \PaxAP\ instances for each of the physics processes under study. 
Each of the \PaxAP\ instances now evolves the combinatorial hypotheses for each event 
according to its process-model, that has been prepared earlier, e.g. with VisualPax.
The virtual method \CClass{finishEvent()} of the \PaxPF\ class then can be used
to process the resulting event hypotheses of all processes with decision techniques such 
as Likelihood methods, Neural Networks, Decision Trees etc. The results of the analysis 
can be written to PAX I/O files or selected observables can be written to a TTree of a ROOT 
file by using the \PaxEITT\ class.

When running over Monte-Carlo data, the generator information can be passed to the PAX factory 
in addition. Furthermore, the \PaxPF\ can be extended by \PaxP\ derivatives exploiting the Monte-Carlo 
truth in order to train the deployed decision techniques in terms of 
ambiguity resolution and background-process suppression. 

VisualPax can be used at any stage of the analysis to re-define process-models or monitor the 
results.

\section{PAX project infrastructure}

The PAX kernel and its officially supported accessories are continuously maintained and 
further developed by currently eight core developers and undergoes regular quality ensurance 
\cite{BorlandArchitect}. 
The PAX webpage \cite{PAXWWW} provides the PAX Users Guide \cite{PAXGuide}, a comprehensive 
text documentation of the PAX toolkit, as well as class reference and fast navigator pages 
for download or online use. 
Version management of the software project is handled with a web-browsable Version Control System (CVS) 
\cite{CVS}\cite{PAXCVS}.

\section{Acknowledgements}
The authors would like to thank Dominic Hirschbuehl, Yves Kemp, Patrick Schemitz, and Thorsten Walter, 
for helpful contributions and feedback.


\end{document}